\documentclass[12pt,preprint]{aastex}
\usepackage{enumerate}

\shorttitle{ Metal-poor HVS candidates }
\shortauthors{ Li et al. }
\begin{document}

\title{Metal-poor hypervelocity star candidates from the Sloan Digital Sky Survey}

\author{Yinbi Li\altaffilmark{1,2}, Ali Luo\altaffilmark{1}, Gang Zhao\altaffilmark{1},
Youjun Lu\altaffilmark{1}, Juanjuan Ren\altaffilmark{1,2}, Fang Zuo\altaffilmark{1}}

\altaffiltext{1}{Key Laboratory of Optical Astronomy, National
Astronomical Observatories, Chinese Academy of Sciences, Beijing
100012, China.}
\altaffiltext{2}{Graduate University of
Chinese Academy of Sciences, 19A Yuquan Road, Beijing 100049, China}
\email{gzhao@nao.cas.cn}

\begin{abstract}

Hypervelocity stars are believed to be ejected out from the Galactic center through dynamical
interactions of (binary) stars with the central massive black hole(s). In this letter, we report
13 metal-poor F-type hypervelocity star candidates selected from 370,000 stars of the
data release 7 of the Sloan Digital Sky Survey. With a detailed analysis of the kinematics of
these stars, we find that seven of them were likely ejected from the Galactic center (GC)
or the Galactic disk, four neither originated from the GC nor the Galactic disk, and the
other two were possibly ejected from either the Galactic disk or other regions. Those candidates
which unlikely originated from the GC or the Galactic disk, may be explained by other mechanisms, like the
tidal disruption of the Milky Way's dwarf galaxies in the Galactic potential, or the gravitational
interactions with a massive black hole at the center of M31 or M32.

\end{abstract}


\section{INTRODUCTION}

The hypervelocity stars (HVSs), recently discovered in the Galactic halo \citep{Brown05,Hirsch05, Edelmann05},
are moving so fast that they may escape from the Galaxy. A natural explanation of these HVSs is that
they are ejected out from the Galactic center (GC) by interactions of stars with the massive black
hole (MBH) or the hypothetical binary MBHs as predicted by \cite{Hills88} and \cite{Yu03}. The ejection mechanisms of HVSs can
be divided into three categories: tidal breakup of binary stars in the vicinity of a single MBH
\citep{Hills88, Yu03, Bromley06}, and the binary stars are probably injected into the vicinity of the MBH from
the young stellar disk in the GC \citep[e.g., ][]{Lu10, Zhang10} or from the Galactic bulge \citep{Perets09a,Perets09b};
single star encounters with a binary MBH \citep{Yu03, Sesana06, Sesana07, Merritt06};
or single star encounters with a cluster of stellar mass black holes around the MBH \citep{Oleary08}.

More than 16 HVSs have been reported in the literature \citep{Brown09}\citep[See also: ][]{Tillich09, Kollmeier09, Tillich10},
most of them are 3--4 M$_{\sun}$ late B-type stars. Assuming a Salpeter initial mass function (IMF),
the expected solar mass HVSs are about 10 times more abundant than 3--4 M$_{\sun}$ HVSs \citep{Brown09}. \cite{Kollmeier09}
systematically searched for such low mass HVSs in about 290,000 spectra of the Sloan Digital Sky Survey (SDSS), however,
they found only 6 metal-poor stars that can be possibly taken as HVS candidates, which might suggest that the IMF of the parent
population of these HVSs is top heavy. A top heavy IMF of the HVS parent population is possibly consistent with the disk
origination \citep{Bartko10, Lu10, Kollmeier10}. In order to distinguish the ejection mechanisms of HVSs and put constraints
on the origin of the parent population of HVSs, it is quite necessary to search for the low-mass HVSs.

In this letter, we aim to find F/G type low mass HVS candidates from the data release 7 (DR7) of SDSS, which provides a
large catalog of stars with precise multi-color photometry and medium resolution ($R\sim 1800$)spectra \citep{York00}.
We report the finding of 13 old metal-poor F-type HVS candidates, and also discuss their possible origins by detailed
analysis of their kinematics. The letter is organized as follows. In Section 2, we introduce the searching process
of our  low-mass HVS candidates. Section 3 discusses their possible
ejection mechanisms through kinematics. Finally, a brief conclusion is given in Section 4.

\section{TARGET SELECTION}

We analyze over 370,000 stars in the SDSS DR7 with their five-band photometry $ugriz$ and spectra,  which are flux- and wavelength-calibrated in $3800-9200$\, {\AA} and are reduced by the automated SEGUE Stellar Parameter Pipeline \citep{Lee08} to produce reliable heliocentric radial velocities RV$_{\sun}$ and atmospheric parameters.

\subsection{High Velocity Objects}

In order to find out the F/G type HVS candidates, we first select the high-velocity
objects from the Galactic radial velocity distribution of F/G type main sequence samples
with the following methods.

\begin{enumerate}

\item Select F/G type main sequence samples with the five photometric criteria and log(g) $>$ 3.0 in
section 2.3.1 of \cite{Ivezic08}: [130775].

\item Find objects at the high velocity end, which significantly deviate from the best fit Gaussian distribution
of line-of-sight velocities in the Galactic rest frame $V_{\rm rf}$: [369].

\end{enumerate}

\begin{figure*}          % FIGURE 1:
\centering
\includegraphics[width=3.25in]{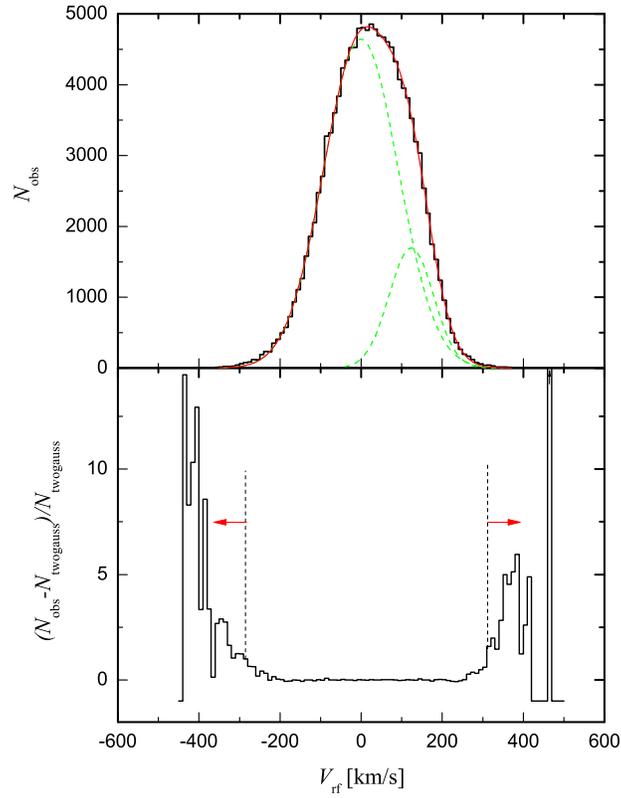}
 \figcaption{ \label{fig:ev1}
 The top panel plots the histogram of line-of-sight velocities in the Galactic rest-frame of 130775 F/G type stars,
 the best-fit two Gaussian function (solid red curve) and the two Gaussian components (dashed green curves). The bottom panel
 is the normalized residuals of the observations from the two Gaussian function. The two red arrows show the region
 of 369 high velocity objects.
}
\end{figure*}

The numbers in brackets indicate the number of stars left after each selection step. The $V_{\rm rf}$
distribution of our 130775 samples, shown in the top panel of Figure \ref{fig:ev1}, can be well fitted by
the two Gaussian distribution (red solid curve). The two Gaussian components have mean velocities
(and scatters) of 0 km s$^{-1}$ (92 km s$^{-1}$) and 124 km s$^{-1}$ (55 km s$^{-1}$) respectively.
Approximately 82\% of the samples belong to the low velocity component, and 18\% of them belong
to the high velocity component (two dashed green curves). These two components probably correspond to
the Galactic halo population (0 $\pm$ 100 km s$^{-1}$) and thick disk population
\citep[90 $\pm$ 45 km s$^{-1}$;][]{Williams11} respectively. The normalized residual of the observation
from this two Gaussian distribution is shown in the bottom panel of the Figure \ref{fig:ev1}, and at its
high velocity tail, high-significance deviations are found. So, we choose the high velocity objects with
$V_{\rm rf} > 309$ km s$^{-1}$ and $V_{\rm rf} < -285$ km s$^{-1}$, which have normalized
residuals larger than 1 and are denoted by two red arrows.

\subsection{HVS Candidates}

To determine which high velocity objects are unbound to the Galaxy, we first obtain the phase space coordinates for them. The Galactic cartesian coordinate system adopted here is centered on the GC: the X axis points from the Sun to the GC with the Sun at $x$ = $-$8 kpc; the Y axis points in the direction of Galactic rotation; the Z axis points towards the Northern Galactic Pole. Assuming  the motion of the local standard of rest (LSR) is 220 km s$^{-1}$, and the velocity of the Sun with respect to the LSR is \citep[11.1 km s$^{-1}$,
12.24 km s$^{-1}$, 7.25 km s$^{-1}$, ][]{Schonrich10}, the phase-space coordinates to the GC for these high velocity objects can be derived by equatorial coordinates ($ra$, $dec$), galactic coordinates ($l$, $b$), $RV_{\sun}$, two components of the proper motion $\mu_{\alpha}\cos(\delta)$, $\mu_{\delta}$, and the heliocentric distances ($D_{\sun}$). Here, $D_{\sun}$ is equal to $\frac{1}{100} \times 10^{0.2~(r - M_{r})}$, where r is the deredden r-band apparent magnitude given by the DR7 and $M_{r}$ is the r-band absolute magnitude calculated by the photometric parallax relation of \cite{Ivezic08}.

Then, we adopt two different Galactic potential models, i.e., a spherically
symmetric model \citep[][hereafter Xue08]{Xue08} and a triaxial model \citep[][hereafter Gnedin05]{Gnedin05} to estimate the
escape velocities ($V_{esc}$). There are totally 13 objects (mpHVS1--mpHVS13, ``mp" means metal-poor) as listed in Table \ref{tab:1}, which are unbound to the Galaxy in the Xue08 model, but only three (mpHVS1--mpHVS3) are unbound in the Gnedin05 model. The rest stars in the 369 high velocity objects are all bound to the Galaxy in both models. We cross-check the $V_{esc}$ for these 13 unbound candidates using other Galactic potential models in the literature, e.g., a spherically symmetric model from \cite{Kenyon08}, two axisymmetric models from \cite{Koposov10} and \cite{Paczynski90} with the same definition of unbound stars as in \cite{Kenyon08}, and find that the $V_{esc}$ from these divergent models are between the values obtained from the Gnedin05 and Xue08 models, and mpHVS1, mpHVS2 and mpHVS3 are unbound to the Galaxy in all models. Note here, a small fraction of the bound stars at the 2-$\sigma$ level could be actually unbound due to the velocity measurement error, but the possibilities are certainly much less than those of unbound candidates. The basic information of these 13 candidates are listed in Table \ref{tab:1}-- \ref{tab:3}. Table \ref{tab:1} summarizes their basic parameters, i.e., the name and its notation, $\mu_{\alpha}\cos(\delta)$ and $\mu_{\delta}$, $RV_{\sun}$, $V_{\rm rf}$, g-band apparent magnitude g, and stellar atmospheric parameters $T_{\rm eff}$, log($g$), and [Fe/H]. Table \ref{tab:2} lists their 6D phase-space coordinates and heliocentric distances $D_{\sun}$. Except mpHVS13, the estimates of $D_{\sun}$ and velocities of other candidates are all valid, because their spectroscopic metallicities [Fe/H] are in the valid range of that given by \cite{Ivezic08}. Table \ref{tab:3} compares their Galactic total velocities $V_{\rm G}$ with $V_{\rm esc}$ of the Gnedin05 and Xue08 models at their Galactocentric distances $D_{\rm G}$, from which we can see whether they are unbound in a certain potential model. In addition, \cite{Kollmeier09} found 6 analogous metal-poor HVS candidates, so we also check whether they are in our high velocity objects. Consequently, only SDSS~J074557.31+181246.7 is selected out and belongs to our bound objects, while the other five candidates are not in our high velocity objects mainly because they are inconsistent with the selection criterion 1 in Section 2.1.

\section{POSSIBLE ORIGINS OF 13 mpHVS CANDIDATES}

\begin{figure*}          % FIGURE 2:
\centering
\includegraphics[width=6.5in]{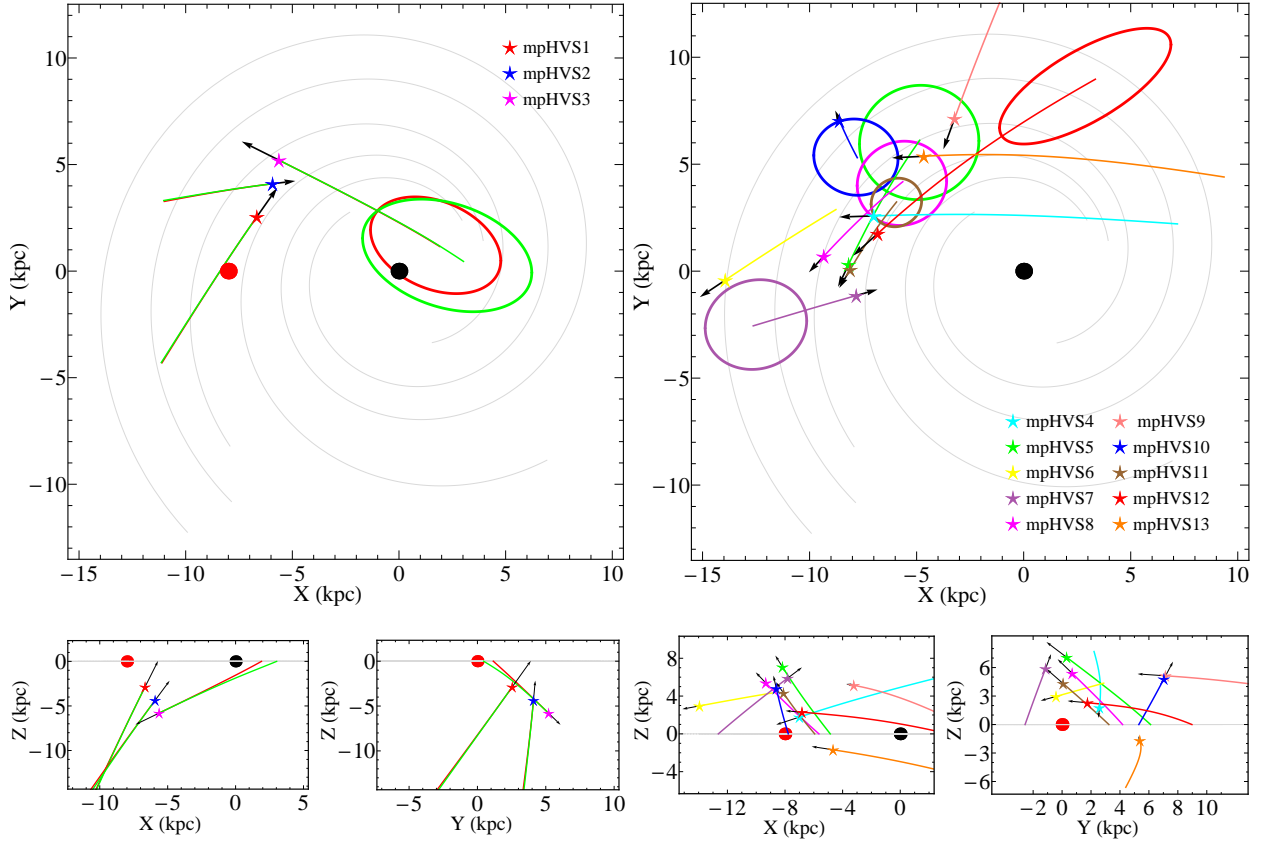}
 \figcaption{ \label{fig:ev2}
  Two-dimensional projections of the orbits of 13 mpHVS candidates in the
Galactic rectangular coordinates. Symbols mark their present positions (star), the GC
(black filled circle) and the Sun (red filled circle). Arrowheads indicate the present
velocities. In the left panel, the red and green curves represent the trajectories of
mpHVS1--3. The red and green ellipses represent the 1-$\sigma$ intersections of the
trajectories of mpHVS3 with the Galactic disk in the potential models of Gnedin05 and
Xue08 respectively. In the right panel, ten curves with different colors respectively
indicate the  trajectories of mpHVS4--13, which are obtained by adopting the Xue08
model. The green, purple, magenta, blue, brown and red ellipses respectively represent
1-$\sigma$ intersections of the trajectories of mpHVS5, mpHVS7, mpHVS8, mpHVS10, mpHVS11
and mpHVS12 with the Galactic disk in the potential model of Xue08.
}

\end{figure*}

To investigate the origins of these 13 mpHVS candidates, it is essential to study their kinematics. By varying the 3D velocities and positions within their measurement errors, we calculate 10000 tracing-back trajectories and the 1 $\sigma$ level intersection places with the Galactic disk for each of mpHVS1--3 with the Gnedin05 and Xue08 models, and those for each of mpHVS4--13 with the Xue08 model, because they are bound in the Gnedin05 model, which are shown in Figure \ref{fig:ev2}. From the left panel of Figure \ref{fig:ev2}, we can see:
(1) the trajectories of mpHVS1 and mpHVS2 do not intersect with the Galactic disk.
(2) the trajectories of mpHVS3 intersect with the Galactic disk about 12 and 15 Myr ago for the Gnedin05 and Xue08 models respectively, and the 1 $\sigma$ intersection places are respectively shown by red and green ellipses.
In the right panel:
(1) the trajectories of mpHVS4 and mpHVS13 do not intersect with the Galactic disk.
(2) the trajectories of mpHVS5, mpHVS7, mpHVS8, mpHVS10, mpHVS11 and mpHVS12 do intersect with the Galactic disk about 10 to 20 Myr ago, and the 1 $\sigma$ intersection regions are depicted by 6 ellipses with different colors.
(3) for mpHVS6 and mpHVS9, most trajectories do intersect with the Galactic disk, but most of the intersection regions are far away from the GC even to hundreds of kpc, so we just describe part of the trajectories for the two candidates, which are calculated with their present 6D phase-space coordinates, and do not show their intersection regions in the right panel.

The trajectories of mpHVS3 are consistent with being ejected out from the GC or the Galactic disk, while those of mpHVS5, mpHVS7, mpHVS8, mpHVS10, mpHVS11 and mpHVS12 are consistent with originating from the Galactic disk (see Figure \ref{fig:ev2}). Therefore, a supernova explosion in a massive tight binary system \citep{Blaauw1961} or dynamical interactions in dense stellar clusters \citep{Poveda1967, Leonard1991, Gvaramadze2009, Gvaramadze2011} are possible originating mechanisms for these seven candidates, and dynamical interactions between (binary) stars and MBH(s) in the GC are also possible ejection mechanisms for mpHVS3. For mpHVS1, mpHVS2, mpHVS4 and mpHVS13, their trajectories are incompatible with the GC origin or the Galactic disk origin mechanisms (see Figure \ref{fig:ev2}). However, their origins may be interpreted as following.

\cite{Sherwin08} proposed that there are a large number of low mass ($\approx$~1 M$_{\sun}$) HVSs ejected by the MBH in the center of M31, and a fraction of them would be moving towards the Milky Way on the solar system side, with large Galactocentric radial approach velocities ($\sim-500$ km s$^{-1}$) exceeding the local escape velocity. The Galactocentric radial velocities of mpHVS1, mpHVS2, mpHVS4 and mpHVS13 are $-476\pm$85 km s$^{-1}$, $-597\pm$177 km s$^{-1}$, $413\pm$96 km s$^{-1}$ and $265\pm$126 km s$^{-1}$ respectively, which means the approaching velocities of mpHVS1 and mpHVS2 are consistent well with the predicted $-500$ km s$^{-1}$, but it seems unlikely that mpHVS4 and mpHVS13 originated from the central MBH of M31. The mpHVS1 and mpHVS2 are low mass stars ($\approx$~1 M$_{\sun}$) and on the side of the Sun (see Fig. \ref{fig:ev2}), considering of these factors, they seem to have been ejected out from the central MBH of M31. The ejection rate of HVSs from M32 is even higher than that from M31 \citep{Lu07}, and thus M32 is also a possible origin of the two mpHVS candidates.

Tidal disruptions of dwarf galaxies in the Milky Way can also produce high velocity stars as proposed by \cite{Abadi09}. And they suggested that the current detected HVSs surrounding the constellation Leo ($l \approx$ 230$^{\circ}$, $b\approx$ 60$^{\circ}$) may be a stream of stars from a dwarf galaxy that was recently tidally disrupted (but see a different explanation of this in \cite{Lu10}), and \cite{Teyssier09} further proposed that such tidal disruption process of the massive satellites or satellites on eccentric orbits can also generate a population of old isolated high speed ``escaped" (unbound) or {\it ``wandering" (bound) stars} \footnote{The ``wandering" stars may possibly be related to the bound stars in the 369 high velocity objects.}. If mpHVS1, mpHVS2, mpHVS4 and mpHVS13 are stars in tidal stripping streams, they may originate from other star streams, as their loci ($l$: 58$^{\circ}$$\sim$69$^{\circ}$; $b$: -45$^{\circ}$$\sim$33$^{\circ}$) are far away from the Leo area. If they are the isolated old ``escaped" tidal stripping stars, we can roughly estimate their stripping times since they only pass though our galaxy once and never come back. Their estimated tidal stripping occurred about $10^{8}$ yr ago, assuming the dwarfs lie at the virial radius. Such a short stripping time suggests that their parent dwarf galaxies or streams may be still detectable along their trajectories. Therefore, further investigations are needed to find such dwarf galaxies, but the current comparatively large data errors impede such studies.

From the results of numerical experiments of mpHVS6 and mpHVS9, they are possibly ejected out from either the Galactic disk by corresponding mechanisms described above, or other places. The Galactocentric radial velocities of mpHVS6 and mpHVS9 are $364\pm$223 km s$^{-1}$ and $-273\pm$262 km s$^{-1}$ respectively, so they are unlikely to be ejected by the central MBH of M31. The tidal interactions between satellite galaxies and our galaxy may be able to explain their origins, but they were surely not ejected out from star streams in the direction of constellation Leo because their loci ($l$:56$^{\circ}$$\sim$185$^{\circ}$; $b$: 26$^{\circ}$$\sim$31$^{\circ}$) are also far away from that region. More exact conclusions about the origins of these two candidates mainly depend on more accurate data measurements in the future.

Note here, the large velocities of these mpHVS candidates could partly be due to orbital motions,
if they were in compact binaries. Below, we give a simple estimation of the effect of binary orbital velocities
on the observed heliocentric radial velocities and the Galactic total velocities. If each of these mpHVS candidates
were in binaries, its companion could be a low mass main-sequence star, a neutron star or a black hole.

\renewcommand{\labelenumi}{\alph{enumi}.}
\begin{enumerate}

\item a low mass ($\leq$ 1 M$_{\sun}$) main-sequence companion: \cite{Rastegaev10}
calculated the distribution of orbital periods for 60 detected metal-poor F,G and early K type binaries, and found the
minimum orbital period was larger than 10 days, and correspondingly the minimum semi-major axis $\geq$ 0.1 AU.
If the mass and semi-major axis of the companion are 1 M$_{\sun}$ and 0.1 AU respectively, the effects due to binary orbital
motion could be most significant. The average effect of the projected
orbital velocities on the line-of-sight is thus about 27 km s$^{-1}$, and the maximum is about 67 km s$^{-1}$,
which introduce only approximate 3 $\sim$ 7 km s$^{-1}$ and 9 $\sim$ 31 km s$^{-1}$ effect to the 3D velocities
of our candidates. In addition, if these candidates are really in such binaries, their heliocentric distances and
Galactic total velocities would be larger than our estimations, and the escape velocities would
be correspondingly smaller, so these binaries should more likely be able to escape from our Galaxy.

\item a neutron star companion: The maximum theoretical mass of a neutron star is $\sim$ 3 M$_{\sun}$ \citep{Kiziltan10},
and if its semi-major is 0.1 AU, the average line-of-sight projected orbital velocity is about 57 km s$^{-1}$, and the
maximum value is about 141 km s$^{-1}$. In this cases, approximate 11 $\sim$ 24 km s$^{-1}$ and 49 $\sim$ 93 km s$^{-1}$
effects on the total space velocities of these mpHVS candidates are introduced.

\item a black hole companion: If the mass and semi-major axis of a black hole are 10 M$_{\sun}$ and 0.1 AU respectively, the average
line-of-sight projected orbital velocity is about 115 km s$^{-1}$, and the maximum velocity is about 284 km s$^{-1}$,
which introduce approximate 35 $\sim$ 69 km s$^{-1}$ and 150 $\sim$ 228 km s$^{-1}$ effects on the total space velocities
of these candidates. In this cases, the binary effect appears significant, so multiple observations are important to investigate whether these HVS candidates are in binaries with black hole companions.

\end{enumerate}

%To close this section, we should also keep in mind all 13 HVS candidates suffer possible large uncertainties in the proper motion estimations, which dominate their 3D total velocities as discussed by \cite{Dong11}. Therefore, we caution here that those 13 candidates, including the highest speed candidates mpHVS1, selected as HVS candidates in this letter could also be statistically bound to the Galaxy. If Dong's model is used to separately estimate their statistic true proper motion at 3~$\sigma$,2~$\sigma$ and 1~$\sigma$, we find that:
%their statistic proper motions are all lower than those given by SDSS, and the 3~$\sigma$ level ones are the lowest. Even so, mpHVS1 is still unbound in both Xue08 and Gnedin05 model, while other candidates could be bound to our Galaxy.

To close the discussion, we note here that only mpHVS1 is unbound with high statistical significance, although all 13 HVS candidates are unbound in some Galactic potential models. The proper motion estimates adopted in our analysis (see Table 1), given by the SDSS,  may suffer some uncertainties as pointed out by Dong et al. (2011), which may effect our estimations on the 3D velocities of those mpHVS candidates.  Applying the model of Dong et al. (2011) (see equation 6), we re-estimate statistically the true  proper motions of those 13 HVS candidates at 3-$\sigma$ significance level. Considering of this correction, we find that the highest velocity  candidate, mpHVS1, is still unbound to the Galaxy in both potential models though with a lower 3D velocity (877 km s$^{-1}$). However, mpHVS4--13 of the other 12 candidates are bound to the Galaxy if adopting the Xue08 potential or mpHVS2--13 are bound if adopting the Gnedin05 potential. So, those mpHVS candidates, except mpHVS1, could be bound halo stars. If they were, it is possible to use them to set constraints on the Galactic potential (W.Brown 2011, private communication).

%(1) the proper motions of these HVS candidates are all decreased, but mpHVS1 is still unbound in both Xue08 and Gnedin05 model even if its 3~$\sigma$ level proper motion is used to evaluate it 3D total velocity.
%(2) mpHVS2 and mpHVS3 are unbound in both models if their 1 or 2~$\sigma$ statistic proper motions are used, and they are also unbound in the Xue08 model if their 3~$\sigma$ statistic proper motion are used.
%(3) mpHVS4 is unbound in Xue08 model if its 1 or 2~$\sigma$ statistic proper motion is used, mpHVS5,mpHVS6,mpHVS7 and mpHVS8 are unbound in Xue08 model using 1~$\sigma$ proper motion, and other candidates are bound to our Galaxy in both models no matter which statistic proper motion is used.

\section{CONCLUSION}

In this letter, we report 13 metal-poor F-type HVS candidates from over
370,000 stars in the SDSS DR7. Through the kinematic analysis, we find that seven of them are likely to have been
ejected from the GC or the Galactic disk. Two of them were possibly ejected out from either the Galactic disk
or other places. Meanwhile, the other four candidates were unlikely to originate from the GC or the
Galactic disk. Those candidates impossibly ejected from the GC or the Galactic disk may be explained by other
mechanisms, i.e., being ejected from the tidal break of the Milky Way's dwarf galaxies \citep{Abadi09, Teyssier09},
or from the center of M31 or M32 by the interactions of stars with the  MBH \citep{Sherwin08}. In order to understand
the origins of these mpHVS candidates well and solve the binary problem, second-epoch observations in future are needed.

\acknowledgements

We thank Yingchun Wei for useful discussions, Jeffrey A. Munn for a comment on the proper
motion provided by the SDSS, \v{Z}eljko Ivezi\'{c} for a comment on evaluating the absolute
magnitude. This work makes use of data products from the SDSS, which is managed by the Astrophysical
Research Consortium for the Participating Institution. The work was funded by the National Science
Foundation of China (NSFC) under grant Nos. 10821061, 10973021, 10973017 and 11103030.

%{\it Facilities:} {MMT (Blue Channel Spectrograph)} {FLWO:1.5m
%(FAST Spectrograph)}

\clearpage
\begin{deluxetable}{lcr  @{$\pm$}  lr  @{$\pm$}  lr  @{$\pm$}  lr  @{$\pm$}  lcccr  @{$\pm$}  l}%\rotate
\centering \tabletypesize{\tiny} \tablewidth{0pt} \tablecolumns{15}
\tablecaption{BASIC PARAMETERS OF HVS CANDIDATES \label{tab:1}}
\tablehead{\colhead{Catalog} & \colhead{Notation} &
\multicolumn{2}{c}{$\mu_{\alpha}\cos(\delta)$} &
\multicolumn{2}{c}{$\mu_{\delta}$} &
             \multicolumn{2}{c}{$RV_{\sun}$} & \multicolumn{2}{c}{$V_{\rm rf}$} &
             \colhead{$g$} & \colhead{${\rm T_{eff}}$} &
             \colhead{log(g)} & \multicolumn{2}{c}{[Fe/H]} \\
           \colhead{} & \colhead{} & \multicolumn{2}{c}{mas/yr} & \multicolumn{2}{c}{mas/yr} &
             \multicolumn{2}{c}{km s$^{-1}$} & \multicolumn{2}{c}{km s$^{-1}$} & \colhead{mag} &
             \colhead{K} & \colhead{} & \multicolumn{2}{c}{}}
\startdata
SDSS~J221853.48$-$004030.9    & mpHVS1   & -41.69&2.88  &  22.69&2.88  &  -432&5  &  -288&5  & 19.32 & 5604 & 3.82  & -1.25&0.10 \\
SDSS~J221625.92+003555.4      & mpHVS2   & -21.12&3.00  &   0.82&3.00  &  -503&9  &  -354&9  & 19.00 & 6307 & 3.85  & -1.44&0.11   \\
SDSS~J222620.74+004135.6      & mpHVS3   &  17.92&3.07  &   3.19&3.07  &   170&8  &   317&8  & 18.86 & 6396 & 3.71  & -1.09&0.06  \\
SDSS~J172915.20+431717.2      & mpHVS4   &   3.52&2.46  &  26.12&2.46  &  -505&5  &  -315&5  & 17.11 & 6559 & 3.83  & -2.00&0.06  \\
SDSS~J125210.66+300147.9      & mpHVS5   & -13.62&2.96  & -12.06&2.96  &   365&8  &   382&8  & 19.31 & 5914 & 4.07  & -1.78&0.05  \\
SDSS J074817.52+360034.4      & mpHVS6   &  -5.16&3.64  & -16.66&3.64  &   335&8  &   313&8  & 19.37 & 6149 & 4.12  & -1.21&0.06  \\
SDSS~J123237.34+165526.9      & mpHVS7   &   7.94&3.04  &  -7.52&3.04  &   396&7  &   359&7  & 18.37 & 6271 & 3.77  & -1.41&0.04  \\
SDSS~J121112.12+402029.5      & mpHVS8   & -14.12&3.22  & -14.78&3.22  &   340&5  &   373&5  & 18.71 & 6216 & 3.84  & -2.00&0.00  \\
SDSS~J172733.94+322459.1      & mpHVS9   &  -9.13&3.30  &  -2.86&3.30  &  -567&18 &  -392&18 & 19.43 & 6308 & 3.81  & -2.00&0.17 \\
SDSS~J171826.16+645745.8      & mpHVS10  & -10.18&3.10  &   0.53&3.10  &   189&8  &   384&8  & 19.28 & 6126 & 3.12  & -1.79&0.00  \\
SDSS~J124814.84+284010.2      & mpHVS11  & -20.48&2.86  & -16.90&2.86  &   361&8  &   371&8  & 18.77 & 5943 & 3.23  & -1.75&0.04  \\
SDSS~J160912.29+345706.8      & mpHVS12  & -35.85&2.44  &   2.32&2.44  &  -435&4  &  -295&4  & 16.90 & 6478 & 3.62  & -2.02&0.03  \\
SDSS J203328.51+142607.0      & mpHVS13  &   5.71&2.81  &   7.27&2.81  &  -492&4  &  -297&4  & 18.37 & 6579 & 3.76  & -2.79&0.10  \\

\enddata
\tablecomments{$\mu_{\alpha}\cos(\delta)$ and $\mu_{\delta}$ are determined by the SDSS and USNO-B
\citep{Gould04, Munn04}, $RV_{\sun}$, ${\rm T_{eff}}$, log(g) and [Fe/H] are obtained by the SEGUE Stellar
Parameter Pipeline, and $V_{\rm rf}$ is the Galactic radial velocity. Here, mp means ``metal-poor".}
\end{deluxetable}

\begin{deluxetable}{lr  @{$\pm$}  lr  @{$\pm$}  lr  @{$\pm$}  lr  @{$\pm$}  lr  @{$\pm$}  lr  @{$\pm$}  lr  @{$\pm$}  l}%\rotate
\centering \tabletypesize{\tiny} \tablewidth{0pt} \tablecolumns{15}
\tablecaption{SPACE POSITIONS AND VELOCITIES OF HVS CANDIDATES
\label{tab:2}} \tablehead{\colhead{Notation} &
\multicolumn{2}{c}{$x$} & \multicolumn{2}{c}{$y$} &
\multicolumn{2}{c}{$z$} &
             \multicolumn{2}{c}{$V_{\rm x}$} & \multicolumn{2}{c}{$V_{\rm y}$} &
             \multicolumn{2}{c}{$V_{\rm z}$} & \multicolumn{2}{c}{$D_{\sun}$} \\
           \colhead{} & \multicolumn{2}{c}{kpc} & \multicolumn{2}{c}{kpc} & \multicolumn{2}{c}{kpc} &
             \multicolumn{2}{c}{km s$^{-1}$} & \multicolumn{2}{c}{km s$^{-1}$} & \multicolumn{2}{c}{km s$^{-1}$} &
            \multicolumn{2}{c}{kpc}}
\startdata

mpHVS1   & -6.7&0.1  &   2.5&0.2  &  -2.9&0.2  &   312&73   &   444&51   &  948&53   &  4.1&0.3     \\
mpHVS2   & -5.9&0.4  &   4.1&0.8  &  -4.4&0.8  &   349&128  &    43&72   &  719&93   &  6.3&1.2       \\
mpHVS3   & -5.6&0.2  &   5.2&0.5  &  -5.8&0.5  &  -569&125  &   285&93   & -421&90   &  8.1&0.7    \\
mpHVS4   & -7.0&0.1  &   2.6&0.3  &   1.8&0.2  &  -526&62   &  -17&29    & -263&33   &  3.3&0.4    \\
mpHVS5   & -8.2&0.0  &   0.3&0.0  &   7.1&0.6  &  -161&107  &  -341&106  &  393&10   &  7.1&0.6    \\
mpHVS6   & -13.9&0.4 &  -0.4&0.0  &   3.0&0.2  &  -388&52   &  -259&119  & -111&104  &  6.7&0.4    \\
mpHVS7   & -7.8&0.0  &  -1.1&0.1  &   5.9&0.7  &   326&90   &    93&89   &  374&19   &  6.0&0.7    \\
mpHVS8   & -9.3&0.1  &   0.7&0.1  &   5.4&0.5  &  -227&89   &  -245&92   &  362&25   &  5.6&0.5    \\
mpHVS9   & -3.2&0.4  &   7.1&0.7  &   5.1&0.5  &  -170&138  &  -469&113  &   47&139  &  10&0.9    \\
mpHVS10  & -8.6&0.1  &   7.0&0.7  &   4.8&0.5  &  -66&125   &   155&75   &  448&109  &  8.5&0.9    \\
mpHVS11  & -8.1&0.0  &   0.1&0.0  &   4.3&0.3  &  -165&65   &  -280&64   &  372&8    &  4.3&0.3    \\
mpHVS12  & -6.8&0.1  &   1.8&0.1  &   2.3&0.2  &  -372&37   &  -336&42   &   47&39   &  3.1&0.3    \\
mpHVS13  & -4.7&0.2  &   5.4&0.3  &  -1.7&0.1  &  -485&76   &  -25&50    &  113&85   &  6.5&0.4    \\

\enddata
\tablecomments{$(x, y, z, V_{x}, V_{y}, V_{z})$ is the 6D Galactic phase-
space coordinate; $D_{\sun}$ is the heliocentric distance.}
\end{deluxetable}

\begin{deluxetable}{c c r  @{$\pm$}  l c r  @{$\pm$}  l c c c c}
\centering
\tabletypesize{\tiny}
\tablewidth{0pt}
\tablecolumns{10}
  \tablecaption{COMPARISON OF THE TOTAL VELOCITIES OF HVS CANDIDATES WITH TWO ESCAPE VELOCITIES OBTAINED BY THE Xue08 and Gnedin05 MODELS\label{tab:3}}

 \tablehead{\colhead{Notation} & \colhead{} & \multicolumn{2}{c}{$D_{\rm G}$} & \colhead{} & \multicolumn{2}{c}{$V_{\rm G}$} &
\colhead{$V_{\rm esc}$ (Xue08)} & \colhead{$V_{\rm esc}$ (Gnedin05)} \\

  \colhead{} & \colhead{} &  \multicolumn{2}{c}{kpc} & \colhead{} & \multicolumn{2}{c}{km s$^{-1}$} &
  \colhead{{ km s$^{-1}$}}& \colhead{{ km s$^{-1}$}}}
\startdata
mpHVS1   & &  7.7&0.3  &  & 1092&103     & 494     &   607        \\
mpHVS2   & &  8.4&1.2  &  &  800&174     & 487     &   599          \\
mpHVS3   & &  9.6&0.7  &  &  763&180     & 477     &   590         \\
mpHVS4   & &  7.7&0.4  &  &  589&76      & 494     &   608         \\
mpHVS5   & & 10.8&0.6  &  &  545&151     & 468     &   583         \\
mpHVS6   & & 14.3&0.4  &  &  479&166     & 447     &   564         \\
mpHVS7   & &  9.9&0.7  &  &  505&128     & 475     &   590         \\
mpHVS8   & & 10.8&0.5  &  &  493&130     & 468     &   584         \\
mpHVS9   & &  9.4&0.9  &  &  501&226     & 479     &   590         \\
mpHVS10  & & 12.1&0.9  &  &  478&182     & 459     &   574         \\
mpHVS11  & &  9.2&0.3  &  &  494&92      & 480     &   595         \\
mpHVS12  & &  7.4&0.3  &  &  503&68      & 496     &   610         \\
mpHVS13  & &  7.3&0.4  &  &  499&124     & 497     &   610         \\

\enddata
\tablecomments{$D_{\rm G}$ and $V_{\rm G}$ are the Galactocentric
distance and 3D total velocity relative to the Galactocentric rest frame respectively. $V_{esc}$
(Xue08)and $V_{esc}$ (Gnedin05) are escape velocities in the Xue08 and
Gnedin05 models.}

\end{deluxetable}

\clearpage

\clearpage
\clearpage

    % REFERENCES
% \bibliographystyle{/home/wbrown/lib/apj} \bibliography{/home/wbrown/text/RefHS}

\begin{thebibliography}{76}
\expandafter\ifx\csname natexlab\endcsname\relax\def\natexlab#1{#1}\fi

\bibitem[Abadi et al.(2009)]{Abadi09}
Abadi, M.\ G., Navarro, J.\ F., \& Steinmetz, M.\ 2009, \apjl, 691, L63

\bibitem[Bartko et al.(2010)]{Bartko10}
Bartko, H., et al.\ 2010, \apj, 708, 834

\bibitem[Blaauw (1961)]{Blaauw1961}
Blaauw, A.\ 1961, Bull. Astron. Inst. Netherlands, 15, 265

\bibitem[Bromley et al.(2006)]{Bromley06}
Bromley, B.\ C., Kenyon, S.\ J., Geller, M.\ J., Barcikowski, E., Brown, W.\ R., \& Kurtz, M.\ J.\ 2006, \apj, 653, 1194

\bibitem[Brown et al.(2009)]{Brown09}
Brown, W.\ R., Geller, M.\ J., Kenyon, S.\ J., \& Bromley, B.\ C.\ 2009, \apj, 690, 1639

\bibitem[Brown et al.(2005)]{Brown05}
Brown, W.\ R., Geller, M.\ J., Kenyon, S.\ J., \& Kurtz, M.\ J.\ 2005, \apj, 622, L33

\bibitem[Dong et al.(2011)]{Dong11}
Dong, R.\ B., et al.\ 2011, \aj, 142, 116D

\bibitem[Edelmann et al.(2005)]{Edelmann05}
Edelmann, H., Napiwotzki, R., Heber, U., Christlieb, N., \& Reimers, D.\ 2005, \apjl, 634, L181

\bibitem[Gnedin et al.(2005)]{Gnedin05}
Gnedin, O.\ Y., Gould, A., Miralda-Escud{\'e}, J., \& Zentner, A.\ R.\ 2005, \apj, 634, 344

\bibitem[Gould \& Kollmeier(2004)]{Gould04}
Gould, A., \& Kollmeier, J.\ A.\ 2004, \apjs, 152, 103

\bibitem[Gvaramadze et al.(2009)]{Gvaramadze2009}
Gvaramadze V.\ V., Gualandris A., \& Portegies Zwart S.\ 2009, \mnras, 396, 570

\bibitem[Gvaramadze \& Gualandris(2011)]{Gvaramadze2011}
Gvaramadze V.\ V. \& Gualandris A.\ 2011, \mnras, 410, 304.

\bibitem[Hills(1988)]{Hills88}
Hills, J.\ G.\ 1988, Nature, 331, 687

\bibitem[Hirsch et al.(2005)]{Hirsch05}
Hirsch, H.\ A., Heber, U., O'Toole, S.\ J., \& Bresolin, F.\ 2005, \aap, 444, L61

\bibitem[Ivezi{\'c} et al.(2008)]{Ivezic08}
Ivezi{\'c}, {\v Z}., et al.\ 2008, \apj, 684, 287

\bibitem[Kenyon et al.(2008)]{Kenyon08}
Kenyon, S.\ J., Bromley, B.\ C., Geller, M.\ J., \& Brown, W.\ R.\ 2008, \apj, 680, 312

\bibitem[Kiziltan et al.(2010)]{Kiziltan10}
Kiziltan, B., Kottas, A., \& Thorsett, S.\ E.\ 2010, astro-ph/10114291v1

\bibitem[Kollmeier et al.(2009)]{Kollmeier09}
Kollmeier, J.\ A., Gould, A., Knapp, G., \& Beers, T.\ C.\ 2009, \apj, 697, 1543

\bibitem[Kollmeier et al.(2010)]{Kollmeier10}
Kollmeier, J.\ A., et al.\ 2010, \apj, 723, 812

\bibitem[Koposov et al.(2010)]{Koposov10}
Koposov, S.\ E., Rix, H.\ W., \& Hogg, D.\ W.\ 2010, \apj, 712, 260

\bibitem[Lee et al.(2008)]{Lee08}
Lee, Y.\ S., et al.\ 2008, \aj, 136, 2022

\bibitem[Leonard(1991)]{Leonard1991}
Leonard, P.\ J.\ T.\ 1991, \aj, 101, 562

\bibitem[Lu et al.(2010)]{Lu10}
Lu, Y., Zhang, F., \& Yu, Q.\ 2010, \apj, 709, 1356

\bibitem[Lu et al.(2007)]{Lu07}
Lu, Y., Yu, Q., \& Lin, D.\ N.\ C.\ 2007, \apj, 666, L89

\bibitem[Merritt(2006)]{Merritt06}
Merritt, D.\ 2006, \apj, 648, 976

\bibitem[Munn et al.(2004)]{Munn04}
Munn, J.\ A., et al.\ 2004, \aj, 127, 3034

\bibitem[O'Leary \& Loeb(2008)]{Oleary08}
O'Leary, R.\ M., \& Loeb, A.\ 2008, \mnras, 383, 86

\bibitem[Paczynski(1990)]{Paczynski90}
Paczynski, B.\ 1990, \apj, 348, 485

\bibitem[Perets(2009a)]{Perets09a}
Perets, H.\ B.\ 2009a, \apj, 690, 795

\bibitem[Perets(2009b)]{Perets09b}
Perets, H.\ B.\ 2009b, \apj, 698, 1330

\bibitem[Poveda et al.(1967)]{Poveda1967}
Poveda, A., Ruiz, J., \& Allen, C.\ 1967, Bol. Obs. Tonantzintla Tacubaya, 4, 86

\bibitem[Rastegaev(2010)]{Rastegaev10}
Rastegaev, D.\ A.\ 2010, \apj, 140, 2013

\bibitem[Sch{\"o}nrich et al.(2010)]{Schonrich10}
Sch{\"o}nrich, R., Binney, J., \& Dehnen, W.\ 2010, \mnras, 403, 1829

\bibitem[Sesana et al.(2007)]{Sesana07}
Sesana, A., Haardt, F., \& Madau, P.\ 2007, \mnras, 379, 45

\bibitem[Sesana et al.(2006)]{Sesana06}
Sesana, A., Haardt, F., \& Madau, P.\ 2006, \apj, 651, 392

\bibitem[Sherwin et al.(2008)]{Sherwin08}
Sherwin, B.\ D., Loeb, A., \& O'Leary, R.\ M.\ 2008, \mnras, 386, 1179

\bibitem[Teyssier et al.(2009)]{Teyssier09}
Teyssier, M., Johnston, K.\ V., \& Shara, M.\ M.\ 2009, \apjl, 707, L22

\bibitem[Tillich et al.(2009)]{Tillich09}
Tillich, A., Przybilla, N., Scholz, R.\ D, \& Heber, U.\ 2009, \aap, 507, L37

\bibitem[Tillich et al.(2010)]{Tillich10}
Tillich, A., Przybilla, N., Scholz, R.\ D, \& Heber, U.\ 2010, \aap, 517, A36

\bibitem[Williams et al.(2011)]{Williams11}
Williams, M.\ E.\ K., et al.\ 2011, \apj, 728, 102

\bibitem[Xue et al.(2008)]{Xue08}
Xue, X.\ X., et al.\ 2008, \apj, 684, 1143

\bibitem[York et al.(2000)]{York00}
York, D.\ G., et al.\ 2000, \aj, 120, 1579

\bibitem[Yu \& Tremaine(2003)]{Yu03}
Yu, Q., \& Tremaine, S.\ 2003, \apj, 599, 1129

\bibitem[Zhang et al.(2010)]{Zhang10}
Zhang, F., Lu, Y., \& Yu, Q.\ 2010, \apj, 722, 1744

\end{thebibliography}

\end{document}